\RequirePackage{rotating}
\PassOptionsToPackage{usenames,dvipsnames}{xcolour}
\documentclass[twocolumn,twocolappendix,iop]{openjournal}

\usepackage{hyperref}
\usepackage{amsmath,amstext}
\usepackage{ae,aecompl}
\usepackage[utf8]{inputenc}
\usepackage[figure,figure*]{hypcap}
\usepackage{url}
\usepackage{mdwlist}
\usepackage{multirow}
\usepackage{booktabs} 
\usepackage{comment}
\usepackage{newtxtext,newtxmath}
\usepackage{xspace}
\usepackage{natbib}
\usepackage[usenames,dvipsnames,svgnames,table]{xcolor}
\hypersetup{colorlinks,linkcolor=blue,citecolor=blue,urlcolor=violet}
\usepackage[a4paper,marginpar=10pt,headheight=20pt, right=1cm, top=1.5in, left=1in, bottom=0pt, footskip=0pt, footnotesep=10pt]{geometry}

\renewcommand{\arraystretch}{1.2}

\usepackage[T1]{fontenc}

\DeclareRobustCommand{\VAN}[3]{#2}
\let\VANthebibliography\thebibliography
\def\thebibliography{\DeclareRobustCommand{\VAN}[3]{##3}\VANthebibliography}

\usepackage{graphicx}	
\usepackage{amsmath}	

\begin{document}

\title{Citizen ASAS-SN Data Release II: Variable Star     Classification Using Citizen Science \vspace{-1.5cm}}

\author{O. Kotrach $^{1*}$}
\author{C. S. Kochanek $^{1,2}$}
\author{C. T. Christy$^{3}$}
\author{T. Jayasinghe$^{4}$ }
\author{K. Z. Stanek$^{1,2}$}
\author{D. M. Rowan$^{1,2}$}
\author{J. L. Prieto $^{5,6}$}
\author{B. J. Shappee $^{7}$}

\affiliation{$^1$Department of Astronomy, The Ohio State University, 140 West 18th Avenue, Columbus, OH 43210, USA }
\affiliation{$^2$Center for Cosmology and Astroparticle Physics,    The Ohio State University, 191 W. Woodruff Avenue, Columbus, OH 43210 }
\affiliation{$^{3}$Department of Astronomy and Steward              Observatory, 933 North Cherry Avenue Tucson, AZ 85721, USA}
\affiliation{$^4$ Independent Researcher, San Jose, California, USA}
\affiliation{$^{5}$Instituto de Estudios Astrof\'isicos, Facultad de Ingenier\'ia y Ciencias, Universidad Diego Portales, Av. Ej\'ercito Libertador 441, Santiago, Chile}
\affiliation{$^6$Millennium Institute of Astrophysics MAS, Nuncio Monse\~nor Sotero Sanz 100, Off. 104, Providencia, Santiago, Chile}
\affiliation{$^{7}$Institute for Astronomy, University of           Hawai’i, 2680 Woodlawn Drive, Honolulu, HI 96822,USA }

\email[$^*$]{kotrach.1@osu.edu}

\begin{abstract}
    We present the second results from Citizen ASAS-SN, a citizen science project for the All-Sky Automated Survey for Supernovae (ASAS-SN) hosted on the Zooniverse platform. Citizen ASAS-SN tasks users with classifying variable stars based on their light curves. We started with 94975 new variable candidates and identified 4432 new variable stars. The users classified the new variables as 841 pulsating variables, 2995 rotational variables, 350 eclipsing binaries, and 246 unknown variables. We found 68\% user agreement for user-classified pulsating variables, 51\% for rotational variables, and 77\% for eclipsing binaries. We investigate user statistics and compare new variables to known variables. We present a sample of variables flagged as interesting or unusual.
\end{abstract}

\vspace{1cm}

\maketitle

\section{Introduction}\label{sec:intro}
    Variable stars are invaluable tools for the study of stellar properties and evolution. Pulsating variables like Cepheids and RR Lyrae stars are used as standard candles, which are essential tools in cosmology \citep[e.g.,][]{Beaton_2018}. Astroseismologists use the pulsations of $\delta$ Scuti stars to probe their interiors \citep{Handler_2009}. Rotational variables allow us to investigate the physics of magnetic dynamos, convection, and angular momentum loss through stellar winds \citep{1981RPPh...44..831M}. Eclipsing binaries provide an accurate and direct way to measure the masses and radii of stars \citep{Torres_2009}. Semi-regular variables provide a precise way of probing galactic structure \citep{Semireg}. Given their utility as astronomical tools, it remains important to find and characterize variable stars.
    
    There are several ongoing wide area variability surveys, including the All Sky Automated Survey for Supernovae (ASAS-SN; \citealt{ 2014ApJ...788...48S, 2017PASP..129j4502K}), the Asteroid Terrestrial-impact Last Alert System (ATLAS; \citealt{2018ApJ...867..105T,2018AJ....156..241H}), the Global Astrometric Interferometer for Astrophysics (Gaia; \citealt{Gaia_overall, gaia}), the Transiting Exoplanet Survey Satellite (TESS; \citealt{TESS}),  and the Zwicky Transient Facility (ZTF; \citealt{Chen_2020}). In good conditions, ASAS-SN images the visible sky nightly to $\sim$18 mag \citep{ 2014ApJ...788...48S, 2017PASP..129j4502K}. Initially, ASAS-SN consisted of two mounts and 8 cameras working in the V-band \citep{2014ApJ...788...48S, 2017PASP..129j4502K}. Analyses of these data led to the discovery of 229,000 variable stars \citep{Jayasinghe_2021}. In 2017 and 2018, ASAS-SN expanded to its current 5 mount/20 telescope configuration and switched to the g-band. A partial analysis based on the new data recovered 263,000 known variable stars, and identified 116,000 new variable stars \citep{Christy_2022}.
    
    In addition to the machine learning methods used for classifications by \cite{Jayasinghe_2018}, \cite{Christy_2022_DR1} used citizen science for variable star classification through the Citizen ASAS-SN project. Citizen ASAS-SN allows users to view and classify light curves into simple 
    categories, namely pulsating, eclipsing, rotational and unknown variables, as well as a "junk" category \citep {Christy_Citizen_Science}. Users can also flag unusual light curves for followup. Data release I of Citizen ASAS-SN found 10420 new variable stars and documented several unusual light curves \citep{Christy_2022_DR1}.
    
    In this paper, we provide a second analysis of Citizen ASAS-SN results. Since the first findings from Citizen ASAS-SN, users examined a series of test samples, flagging 4432 new variables and recovering 56364 variables. In Section \S2, we outline our analysis and provide updated statistics on user performance and activity in Citizen ASAS-SN. We also discuss both the new and previously cataloged variables. In Section \S3, we examine the interesting objects that users tagged and discussed through the tagging and chat board features. In Section \S4, we summarize our findings.
\vspace{0.5cm}

\section{UPDATED PROJECT DESCRIPTION AND RESULTS}
    Citizen ASAS-SN is hosted by the Zooniverse\footnote{https://www.zooniverse.org/} platform, allowing users to access light curves and vote for a classification \citep{Christy_Citizen_Science}. For a given variable, users are shown three versions of the light curve: one phased by the best Lomb-Scargle period, one phased by twice the best Lomb-Scargle period \citep{GLS1, GlS2}, and the observed light curve, as long period and rotational variables are sometimes more identifiable from their observed light curves. We show twice the GLS period because it is not uncommon for the GLS period of an eclipsing binary to be $1/2$ the true period. 
     
     Users are given an explanatory tutorial on how Citizen ASAS-SN works, including example light curves of each of the possible classifications. The 5 options users are presented with are pulsating variable, eclipsing binary, rotational variable, unknown variable, and junk, with unknown variables being chosen for ambiguous light curves. This tutorial can be accessed again at any time. The users are then tasked with classifying a training sample of light curves with feedback on their choices. A light curve is considered "retired" once it has either received 10 total classifications across all 5 options, or it has received 5 junk classifications. 
     
    \begin{figure}
        \centering
        \includegraphics[width=8.3cm]{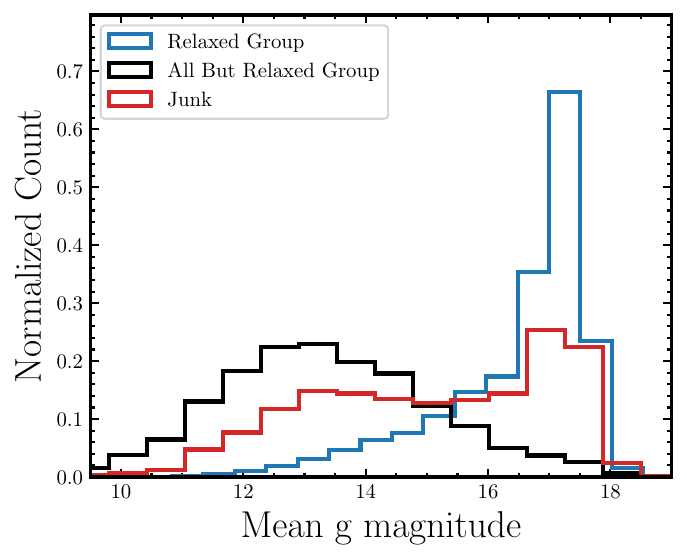}
        \caption{The normalized g magnitude distribution of the Relaxed Group variables, all variables excluding the Relaxed Group, and variables classified as junk by users.}
        \label{fig:g mag}
    \end{figure}

\subsection{GROUP DESCRIPTIONS}
In the first data release, users classified variables with $\delta < -\rm60^{\circ}$ and $g < 18 $ mag, with various cuts in light curve and GLS periodogram statistics \citep{Christy_2022_DR1}. Since then, there have been several new groups of light curves that users have been tasked with classifying, which we will call the Relaxed Group, the Standard Group, the No Periodogram (NP) Group, and the Machine Learning Group. The Relaxed Group includes light curves selected by relaxing the cuts on string length statistics\footnote{The Lafler-Kinmann (REF) string length statistic compares the dispersion of adjacent points of a light curve to the root mean square variance of the light curve \citep{Lafler}.} and decreasing the required peak power in the GLS periodogram compared to \cite{Christy_2022_DR1}. Consequently, the Relaxed Group includes a large number of fainter variables, as seen in the g-band magnitude distribution shown in Figure \ref{fig:g mag}. This is also where we see a peak in the sources classified as junk. This group was chosen to investigate how users would classify noisier light curves. The Standard Group variables were selected with the same criteria as \cite{Christy_2022_DR1}, but with  $-60^{\circ}\leq \delta < -54^{\circ}$. The NP  variables were selected purely based on photometery with declinations of $-54^{\circ}\leq \delta < -50^{\circ}$. This was done to investigate variable selection without the computational expense of running periodograms for all stars. The Machine Learning Group candidates were selected as variables using the updated RF classifier from \cite{Christy_2022_DR1} in order to  evaluate the performance of both the RF classifier and the users. Since all the Machine Learning Group stars were identified as variables by the RF classifier, there should be no new variables, as we see in Table \ref{Table: Group}. These variables were chosen from the northern hemisphere. This group has a 5\% junk classification rate, leading to the question of whether the users or the RF classifier is incorrect. We explore this later in section \S  2.4. More statistics on each group are shown in Table \ref{Table: Group}.

    \begin{figure}
        \centering
        \includegraphics[width=8.4cm]{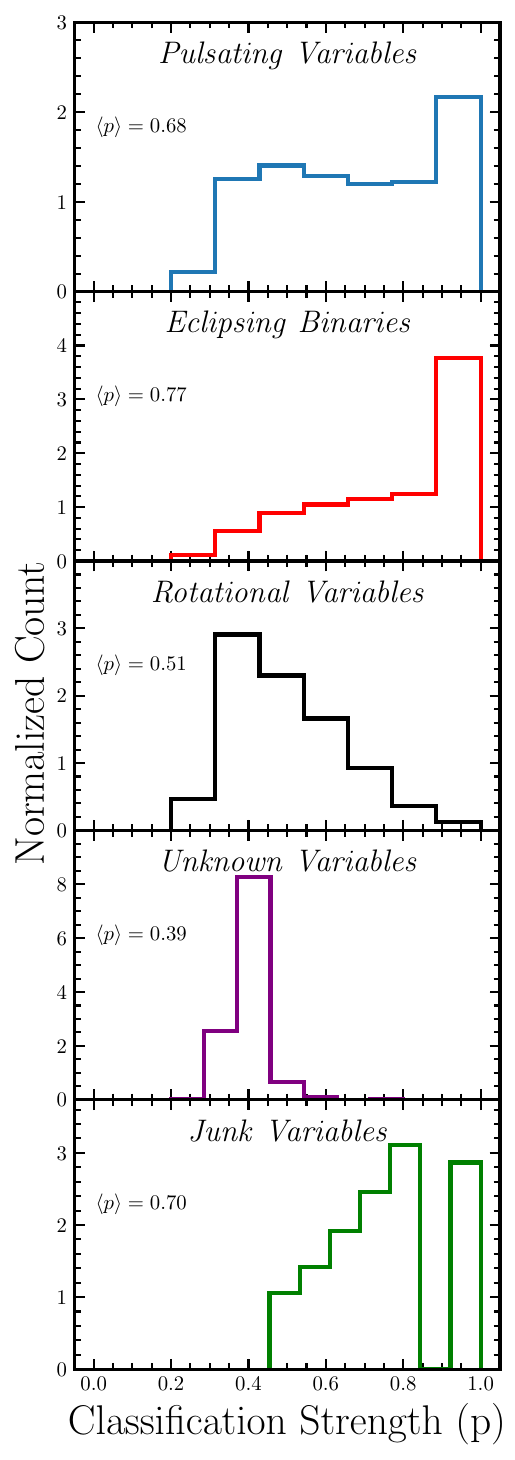}
        \caption{The normalized distributions of classification strengths for the different classification options for all retired variables. }
        \label{fig:All Class}
    \end{figure}

 \begin{figure*}[]
        \centering
        \includegraphics[width=14.9cm]{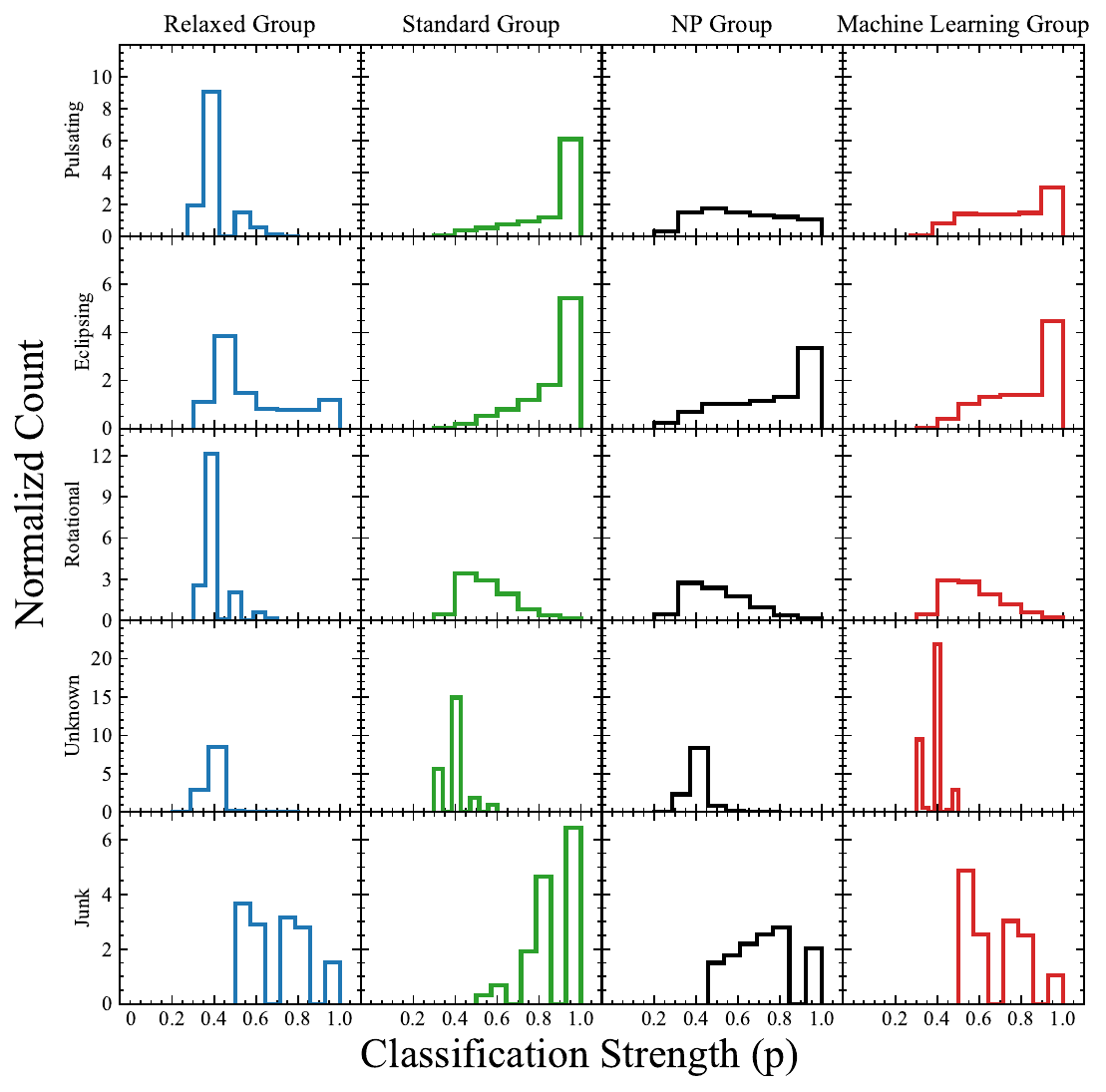}
        \caption{The normalized distribution of classification strengths for the different classification options and the different groups for all retired variables.}
        \label{fig:Batch class}
    \end{figure*}
\newpage

    \begin{figure*}
    \centering
    \includegraphics[width=14.9cm]{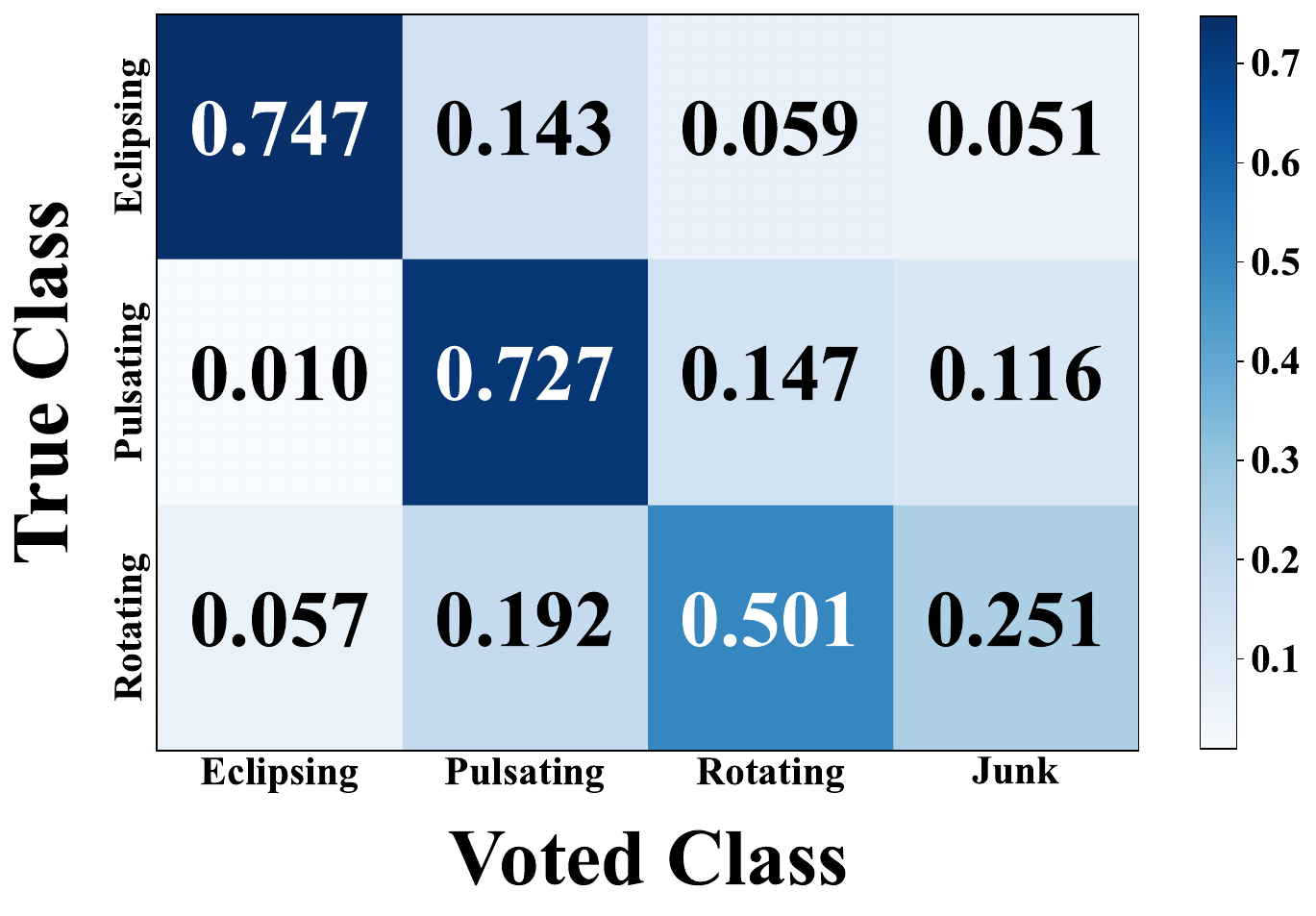}
    \caption{The normalized confusion matrix between user classifications (voted class) and true classifications (true class) from the ASAS-SN, VSX, OGLE, and Gaia catalogs.}
    \label{Knwon Confusion Matrix}
\end{figure*}

\setlength{\tabcolsep}{4.5pt}
\renewcommand{\arraystretch}{2 }
\begin{table*}
    \centering
    \caption{Breakdown of the composition of each group.}
    \begin{tabular}{lcccc}
    \toprule
    \textbf{} & \textbf{Relaxed Group} & \textbf{Standard Group} & \textbf{No Periodogram Group} & \textbf{Machine Learning Group} \\
    \midrule
     Declination   & $\delta<-\rm60^{\circ}$   & $-60^{\circ}<\delta<-54^{\circ}$ & $-54^{\circ}<\delta<-50^{\circ}$ & $10^{\circ}<\delta<88^{\circ}$  \\

     Mean mag & 16.47 & 12.68 & 13.13 & 12.98 \\

     N Total & 19098 & 15758 & 36412 & 25000 \\

     N Retired & 19097 & 15708 & 35213 & 24957 \\

     N Recovered & 1246 & 5036 & 25125  & 24957 \\

     N Total New & 746 & 241 & 3445 & --- \\

     N Eclipsing (New) & 876 (182) & 1922 (58) & 2344 (110) & 10028  \\

     N Pulsating (New & 1033 (190) & 1772 (115) & 10260 (536) & 11781  \\

     N Rotating (New) & 782 (301) & 270 (61) & 10685 (2633) & 1717  \\

     N Unknown (New) & 314 (73) & 25 (7) & 962 (166) & 133  \\

     N Junk & 16092 & 11719 & 10962 & 1298 \\
     \bottomrule
    \end{tabular}
    \label{Table: Group}
\end{table*}

\begin{figure}
    \centering
    \includegraphics[width=8cm]{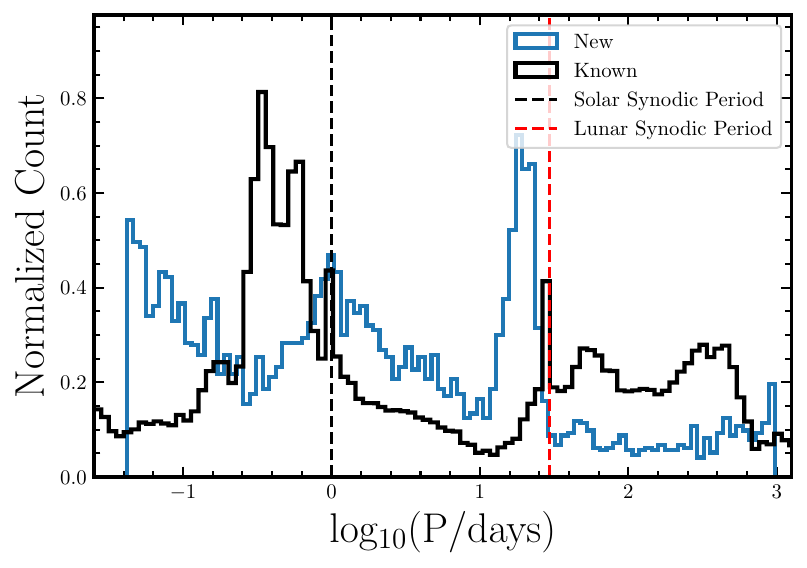}
    \caption{The normalized frequency distribution of the periods at which new and known variables were discovered/recovered. Note the sharp period spikes in the known variables at the lunar and solar synodic periods.}
    \label{fig:period-freq}
\end{figure}

\begin{figure*}
    \centering
    \includegraphics[width=15cm]{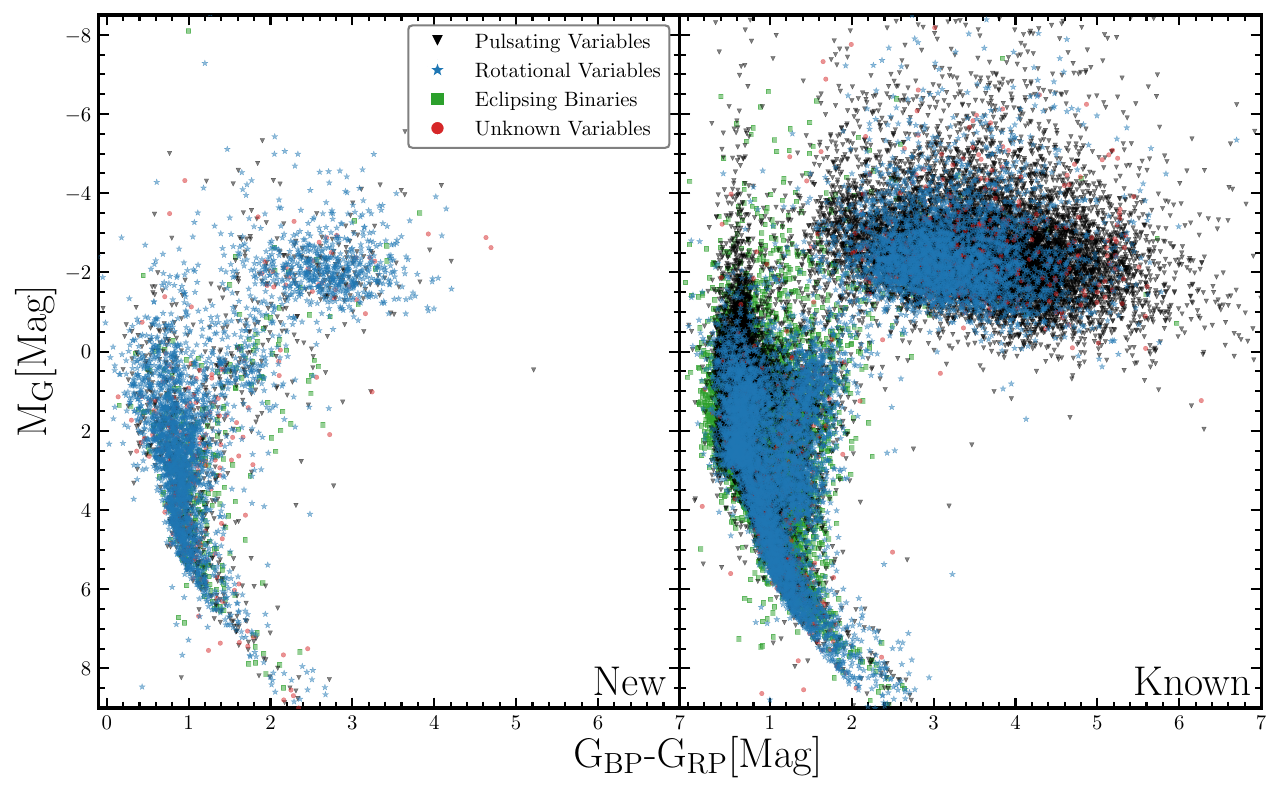}
     \caption{The extinction-corrected Gaia color magnitude diagrams for both new (left) and known variables (right). Classifications are the Citizen ASAS-SN voted types.}
    \label{fig:color}
\end{figure*}

\begin{figure*}
    \centering
    \includegraphics[width=15cm]{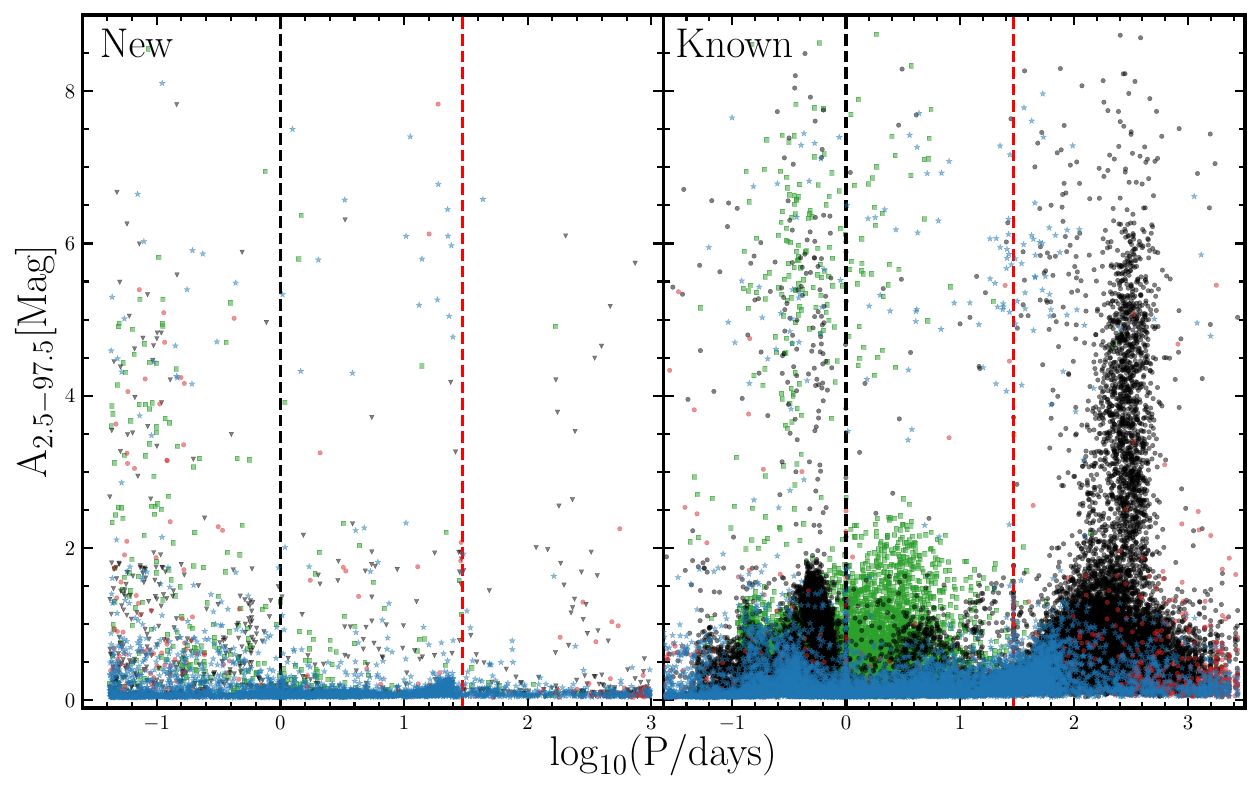}
    \caption{The magnitude variability amplitude versus period plot for new variables (left) and known variables (right). Black points correspond to pulsating variables, blue points to rotational variables, green to eclipsing binaries, and red to unknown variables. The dashed black (red) line lies at the solar (lunar) synodic period. Classifications of variables are assigned by Citizen ASAS-SN voted types.}
    \label{fig:PeriodLum}
\end{figure*}

\setlength{\tabcolsep}{4.5pt}
\renewcommand{\arraystretch}{2.5}
\begin{table*}
    \centering
    \caption{Breakdown of recovered variables from the ASAS-SN, VSX, Gaia VariSummary, and OGLE catalogs. Variable types are given by the user voted classification. N Overlap is the number of variables in a catalog that have been found in at least one other catalog.}
    \begin{tabular}{rrrrrrrr}
    \toprule
    \textbf{} & \textbf{Total} & \textbf{Eclipsing} & \textbf{Pulsating} & \textbf{Rotational} & \textbf{Unknown} & \textbf{Junk} & \textbf{N Overlap} \\
    \midrule
    N Candidates & 94975 & 24846 & 15170 & 13454 & 1434 & 40071 & ---  \\
    ASAS-SN & 35829 & 18299 & 10851 & 4513 & 291 & 1875 & 35829\\
    In VSX & 45713 & 21670 & 13891 & 6344 & 444 & 3364 & 45669 \\
    In Gaia & 49826 & 21380 & 13324 & 7164 & 571 & 7387 & 39488 \\
    In OGLE & 515 & 422 & 13 & 21 & 3 & 56 & 505 \\
    Recovered & 56364 & 14410 & 23006 & 9516 & 731 & 8701 & ---\\
    New & 4432 & 841 & 350 & 2995 & 246 & --- & ---\\
    \bottomrule
    \label{Catalog Table}
    \end{tabular}
\end{table*}

\subsection{USER STATISTICS AND PERFORMANCE}
    For the second data release of Citizen ASAS-SN, 7485 users contributed 1665695 classifications of 244599 light curves. The vast majority of these classifications were of the variable candidates, with only 870 being of the training set variables. Most of the users contributing to the second data release had already classified the training set from the first data-release. Of the 7458 participants, 2338 had unregistered accounts, and 5147 had registered accounts. The registered users were responsible for almost all (94\%) of the classifications.

\begin{figure*}
    \centering
    \includegraphics[width=15cm]{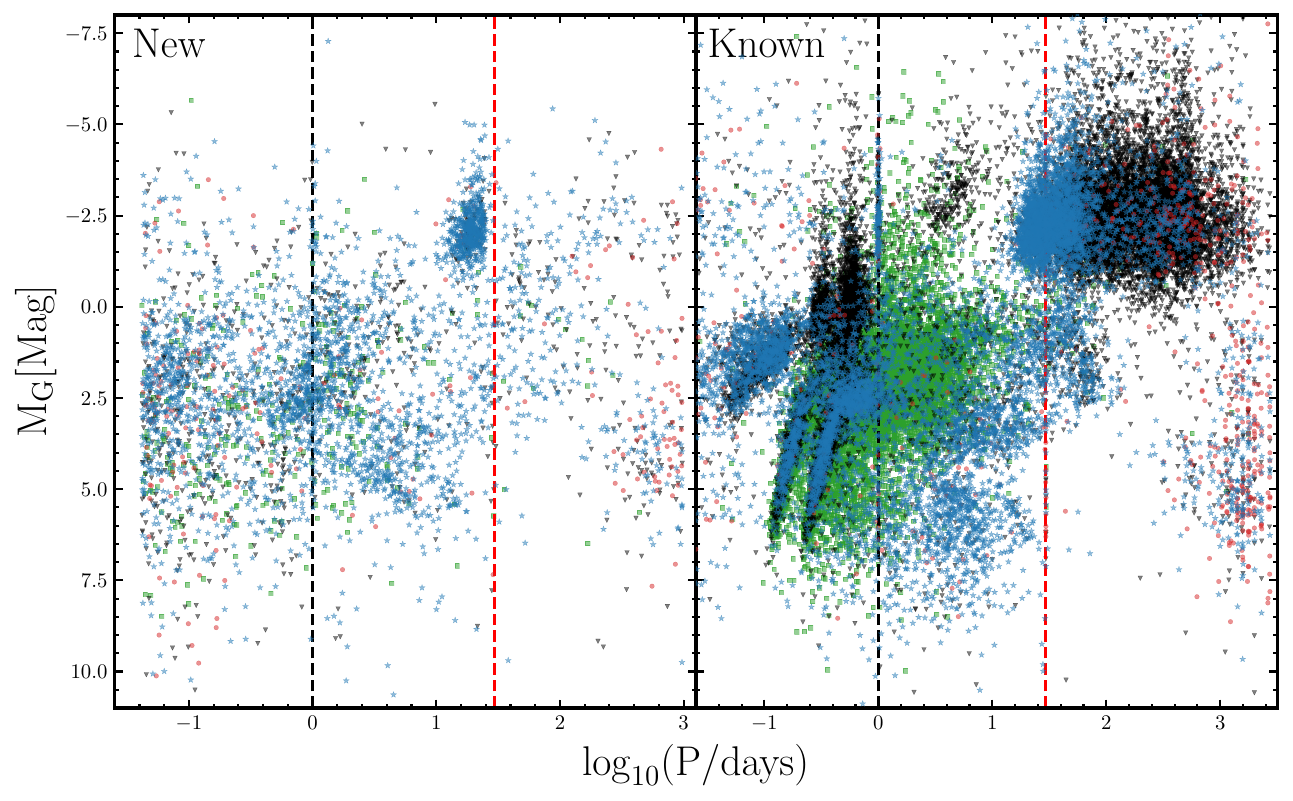}
    \caption{The absolute G magnitude versus period distributions for new variables (left) and known variables (right). Black points correspond to pulsating variables, blue points to rotational variables, green to eclipsing binaries, and red to unknown variables. The dashed black (red) line lies at the solar (lunar) synodic period. Classifications of variables are assigned by Citizen ASAS-SN voted types.}
    \label{fig:MagnitudePeriod}
\end{figure*}

    We define the classification strength $p$ as the ratio between the number of votes for a classification and the total number of votes. It is a measure of how much user agreement exists for a variable, with a maximum of $p = 1$ if all votes are for the same classification and a minimum of $p = 0.2$ for votes equally spread out over all 5 classifications. Figure \ref{fig:All Class} shows the distributions of classification strengths for all retired variable candidates. Mean classification strengths were highest for eclipsing binaries, junk, and pulsating variables, with <$p$> = 0.77, 0.70, and 0.68 respectively. There is increased disagreement between users for rotational variables and unknown variables, with <$p$> = 0.51 and 0.39, respectively. The mean classification strengths for eclipsing binaries, pulsating variables, unknown variables, and rotational variables are relatively unchanged from the first data release \citep{Christy_2022_DR1}. The mean classification strength for junk has decreased by 0.06 from \cite{Christy_2022_DR1}.

    When we break down the classification strength distributions by group, as shown in Figure \ref{fig:Batch class}, we see that the Relaxed group's classification strength distributions are shifted to lower values for all classes. This is likely due to the large number of fainter variables and the weaker selection criteria in the Relaxed group compared to the other groups. For the other samples, the classification strength distributions peak at higher values of $p$ in the pulsating, eclipsing, and junk classes, indicating increased user agreement in classifications in these classes and groups. In the Standard, NP, and Machine Learning Groups, we see similar distributions for rotational and eclipsing variables. The classification agreement does not vary much between groups for unknown variables.

\begin{figure*}[htp]
    \centering
    \includegraphics[scale=0.45]{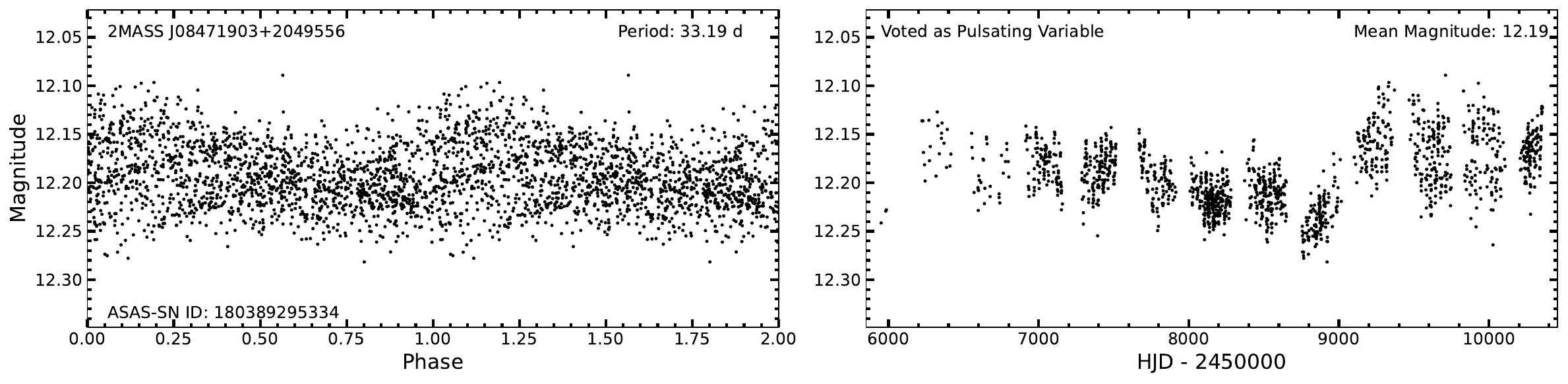}
    \includegraphics[scale=0.45]{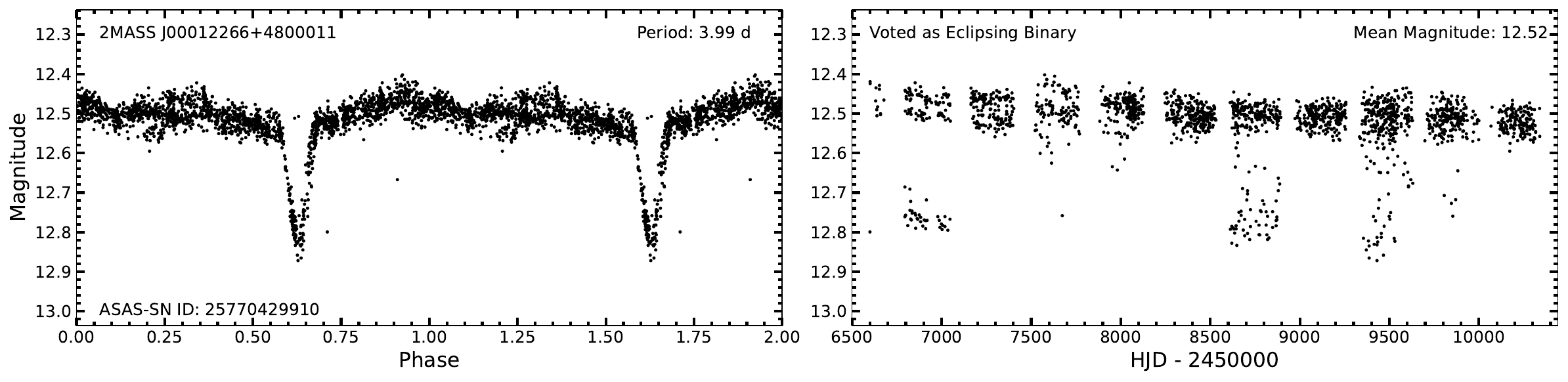}
    \includegraphics[scale=0.45]{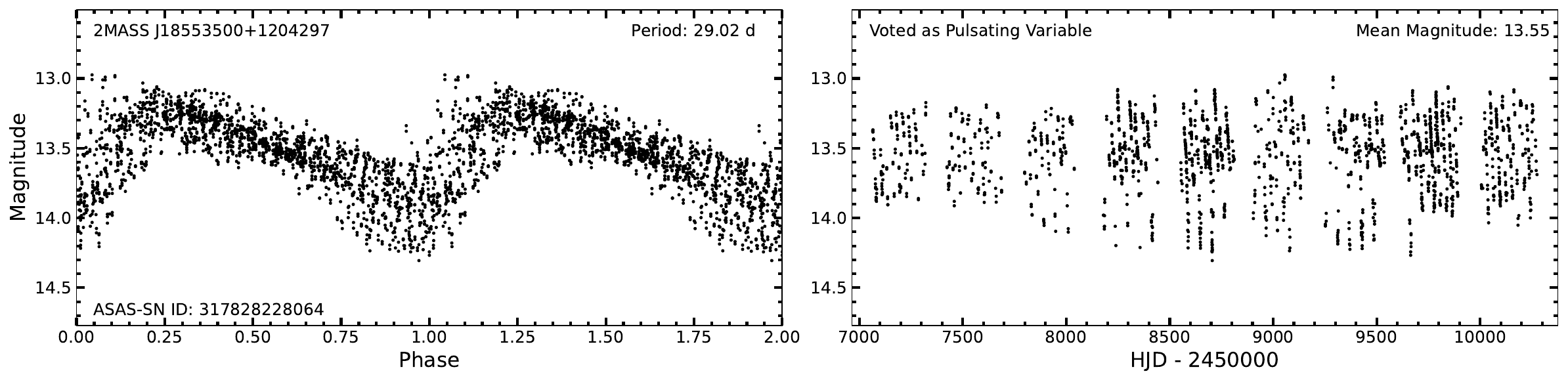}

    \caption{Some of the interesting variables, all with phase folded light curves (left column) and the observed light curve (right column).}
    \label{fig:interesting_vars_1}
\end{figure*}

\begin{figure*}[htp]
    \centering
    \includegraphics[scale=0.45]{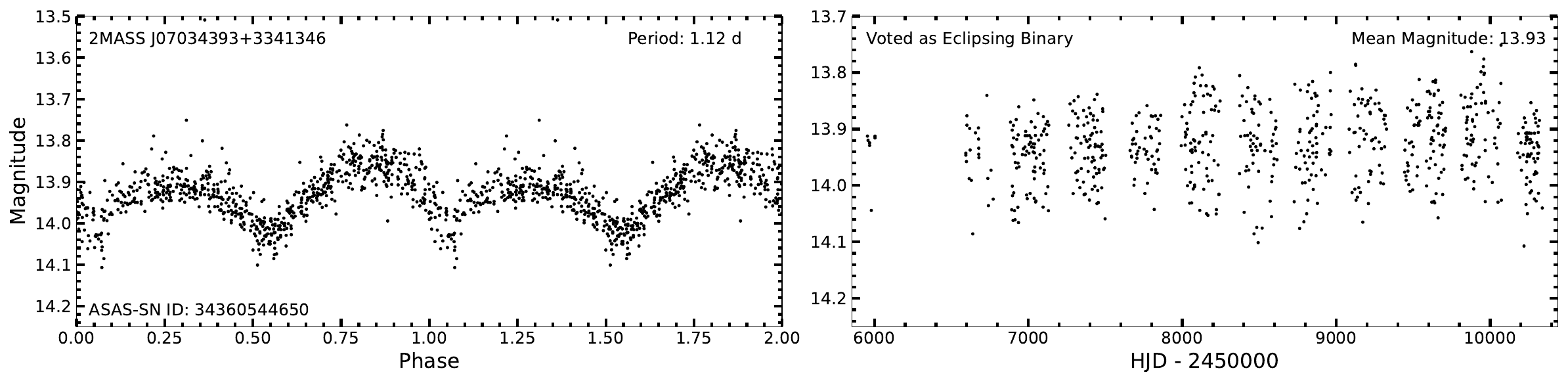}
    \includegraphics[scale=0.45]{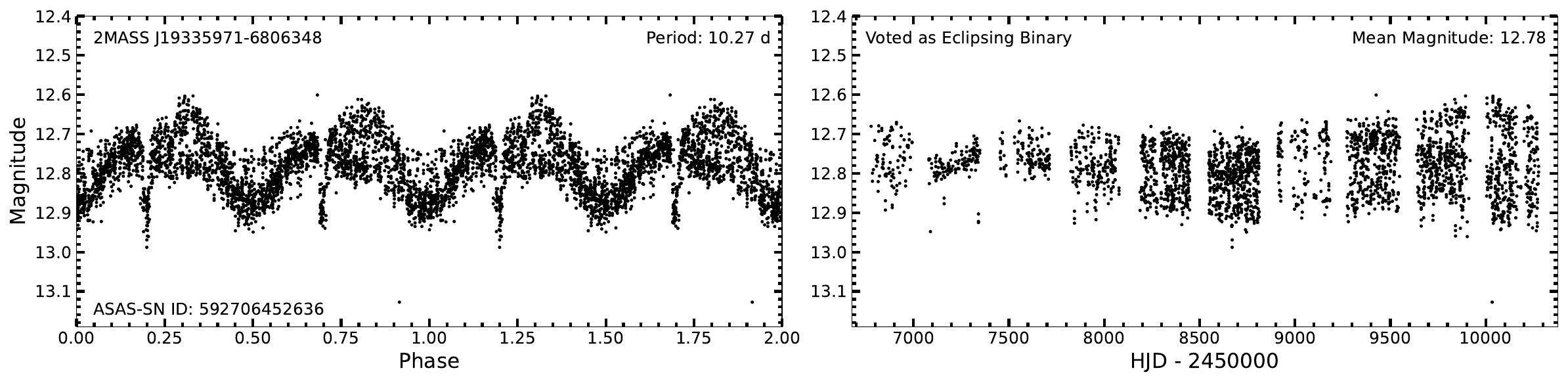}
    \includegraphics[scale=0.45]{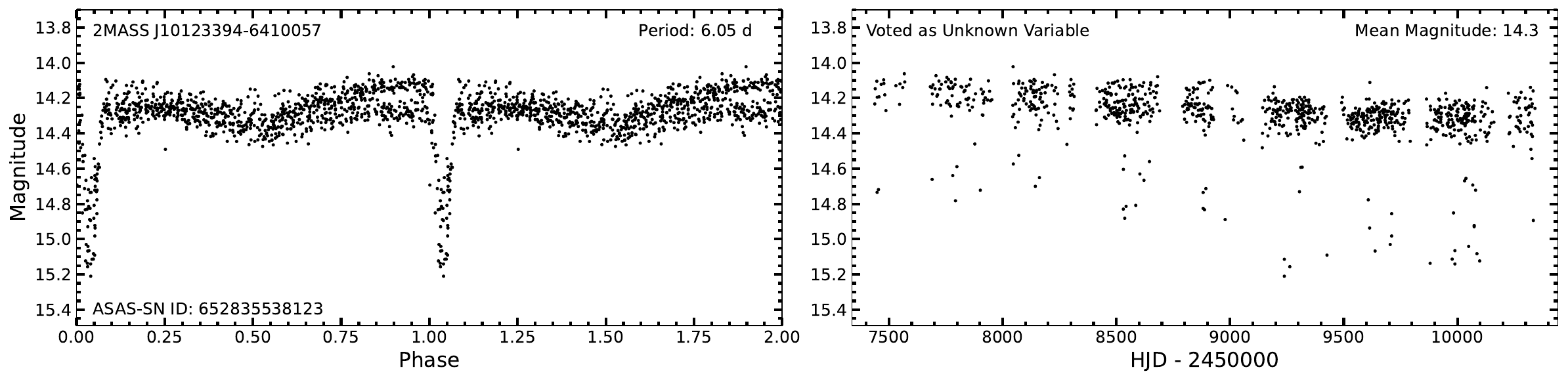}
    \includegraphics[scale=0.45]{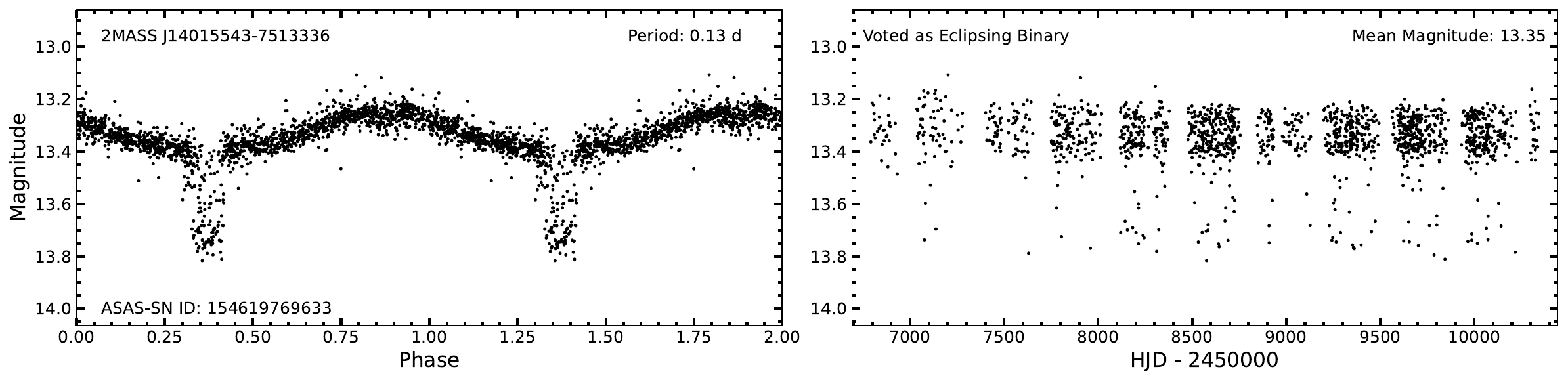}
    \includegraphics[scale=0.45]{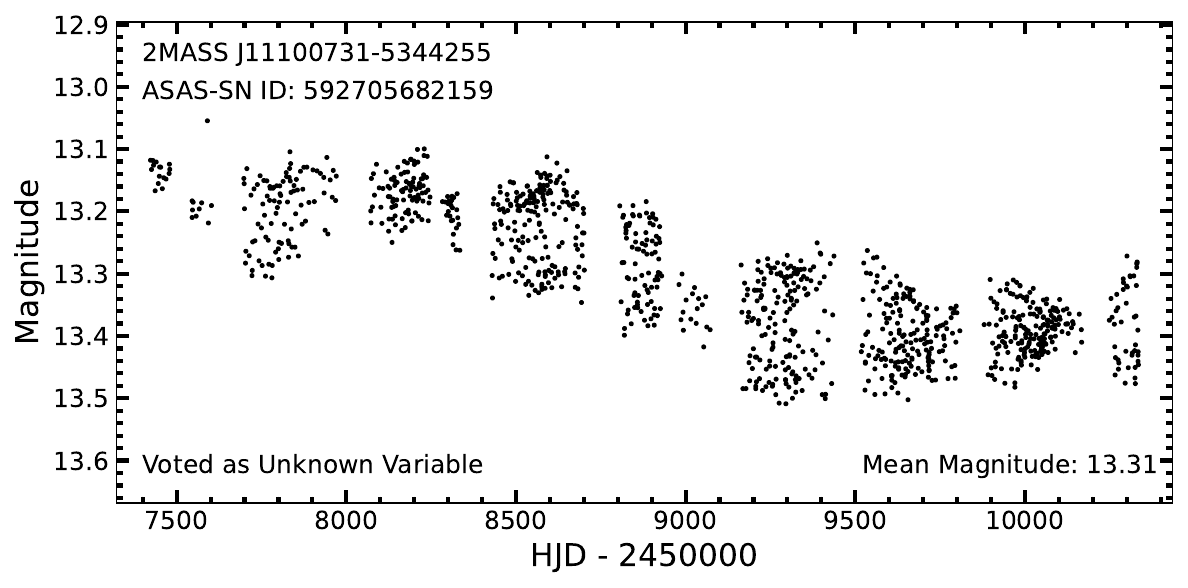}
    \caption{The remaining interesting variables. 2MASS J11100731-5344255's phase folded light curve is not shown.}
    \label{fig:interesting_vars_2}
\end{figure*}

\subsection{NEW AND RECOVERED VARIABLES}
    Of the 94975 retired variable candidates, 56364 variables were recovered from cross-matches to AAVSO VSX \citep{VSX}, ASAS-SN \citep{2014ApJ...788...48S, 2017PASP..129j4502K, JayaASASSN}, Gaia \citep{Gaia_overall, gaia_varisummary} and OGLE \citep{ogle3, ogle4}. Using a matching radius of 16 arcsec (the ASAS-SN FWHM), we found 35829 matches to the ASAS-SN catalog, 45713 matches in VSX, 49826 matches in Gaia's DR3 VariSummary, and 515 matches in OGLE. A breakdown of the recovered variables is shown in Table \ref{Catalog Table}.

    We can evaluate user accuracy by comparing their classifications to cataloged classifications, as shown in Figure \ref{Knwon Confusion Matrix}. We find that users accurately identified 75\% of eclipsing binaries, 73\% of pulsating variables, and 50\% of rotational variables. Users incorrectly identified true rotational variables as junk about 25\% of the time. Users noticeably classified more recovered variables as junk than in \cite{Christy_2022_DR1}, especially variables that are true rotational and pulsating variables. Users classified 1875 variables as junk in the ASAS-SN catalog, 3364 in VSX, 7387 in Gaia, and 56 in OGLE. However, we take the cataloged classifications to be correct.

    Once recovered variables were removed from our sample, we placed cuts to remove solar and lunar aliases, removing variables with periods between 0.96 to 1.03 days (solar alias) and 29.17 to 29.92 days (lunar alias). We also removed any variables with a period less than 1 hour or greater than 1000 days, as described in \citep{Christy_2022_DR1}.  Lastly, we removed variables that had periods between 0.25 and 0.51 days while also having mean g magnitudes larger than 12. This was done to eliminate a clump of variables that were clearly junk once we examined their light curves. After dropping light curves classified as junk, we were left with a sample of 4432 new variable stars. As classified by our users, we found 841 pulsating variables, 2995 rotational variables, 350 eclipsing binaries, and 246 unknown variables. 746 of the new variables come from the Relaxed group, 241 come from the Standard Group, and the remaining 3445 come from the NP Group.

    In Figure \ref{fig:period-freq}, we compare the period distributions of the new and known variables. Many of the new variables have short periods (P $\lesssim$ 29 days). The g band light curves used for Citizen ASAS-SN should be significantly better at finding such short period variable stars because the longitude spread of the four ASAS-SN sites significantly improves the sampling of shorter time scales. We also see that the cataloged known variables are pretty clearly contaminated by solar and lunar aliases. In Figure \ref{fig:color}, we show the Gaia $M_G$ and $G_{BP} - G_{RP}$ color-magnitude diagram colored by user assigned classifications. We used distances from \cite{gaia} and corrected for extinction using the 3D extinction model of \cite{dust}, which is based on \cite{Drimmel}, \cite{Marshall}, and \cite{Green}. The new variables are more concentrated on the main sequence, consistent with the shorter periods: 80.5\% of the new variables are located on the main sequence, while only 64.8\% of the known variables are. In Figure \ref{fig:PeriodLum} we show the amplitudes as a function of period, where the amplitude $A_{2.5-97.5}$ is the magnitude range spanning 2.5\% to 97.5\% of the light curve. Variables are colored by user assigned classifications. The new variables are concentrated at lower amplitudes, as one might expect. Figure \ref{fig:MagnitudePeriod} shows the absolute G magnitude as a function of period colored by user assigned classifications. New variables are concentrated at lower periods and luminosities, which makes sense given the improved performance at short periods.
    \\

\subsection{USER CLASSIFICATION TESTS AND GROUP RESULTS}
    
    As mentioned in Section \S2.1, each group had different reasons for being created. The Relaxed group was designed to investigate user classifications of nosier light curves, the Standard Group consisted of variables selected with the same criteria from \cite{Christy_2022_DR1}, but at a higher declination range, the NP Group was designed to investigate user classifications of light curves selected without computing periodograms, and the Machine Learning Group was designed to evaluate the purity of the training set used in \cite{Christy_2022_DR1} and \cite{Jayasinghe_ML}. The lesson from the Relaxed group is that we cannot relax the selection criteria without getting a large number of junk variables (84\% in Table \ref{Table: Group}). Since this is also the same declination that was surveyed in \cite{Christy_2022_DR1}, it is also reasonable to assume most of the real variables were found in \cite{Christy_2022_DR1}. The junk and unknown classification rates for the NP Group ($\sim$ 31\% and 2.7\%) are comparable to the Standard Group, which means we can effectively identify variable candidates before doing the computationally expensive periodograms. Of the Machine Learning Group variables defined as real by the RF classifier in \cite{Jayasinghe_ML}, the users flagged 5\% of them as junk. When we investigated these light curves, we found that roughly half were clearly variables, so the machine learning selection had a false classification rate of only 2-3\%. We also examined the variables removed for periods that are likely solar or lunar aliases and found that the users classified $\sim$ 95\% of these as junk, which we verified by randomly inspecting a subset of them.

\section{INTERESTING VARIABLES}
   The users can also engage in discussions on the Citizen ASAS-SN talk forum regarding light curves they find interesting. These discussions totaled 125,006 words (roughly 250 pages of text) of 10955 light curves. We investigated the light curves with the largest number of words written about them, since more words likely indicates a more interesting variable. Most of the examples were due to an under appreciation of the semi-regular variability of rotational and (most) giant variables. Users also tagged variables for exhibiting behaviors corresponding to more than one classification (e.g., eclipsing binary and rotational variable). When we randomly investigated a subset of these variables we found no real examples of such behaviors. However we did find a few such systems through the heavily discussed light curves.
 
    Figures \ref{fig:interesting_vars_1} and \ref{fig:interesting_vars_2} show the eight phased and un-phased light curves of these stars. Variables were phased by the highest power GLS periodogram period. All eight of these variables have been previously cataloged. Catalog classifications were generally not in agreement with each other, or with the user classifications.  

    2MASS~08471903$+$2049556 is a long period variable (LPV) identified by \cite{Pepper} and as a rotational variable in ASAS-SN. The light curve is not remarkable for an LPV, and it is included here as an illustration of how irregular variability was a difficult problem for our users.

    2MASS~00012266$+$4800011 is classified as an EB eclipsing binary with a period of 1.3316 days in the VSX catalog. \cite{Jaya2019} found the same classification and period. It was classified as an eclipsing binary by the users and as an RR Lyra by Gaia. Here we find a longer period of 3.99 days, but the curious property of this source lies in the un-phased light curve, in which only some seasons show the apparent eclipses. Our GLS periodogram also recovers the 1.33 day peak, though with a smaller power. After running a BLS periodogram, this star appears to be an eclipsing binary with a $\sim 4$ day period with some out of eclipse rotational variability at a period $\sim 1.3$ days. There are other examples of such transient eclipsing binaries such as SS~Lac (see, e.g., \citealt{Torres2000}) and V907~Sco (see \citealt{Lacy1999}) as well as in the OGLE samples of eclipsing binaries (\citealt{Graczyk2011}). Note that SS~Lac and V907~Sco are triples, so dynamical effects from the tertiary drive the changing eclipse geometry.

    2MASS~18553500+1204297 (V489 Aql) is a well-known RV~Tauri, Type~II Cepheid variable found by \cite{Ceraski1905}. It was classified as a pulsating variable by users and a Type II Cepheid (RV Tauri) by OGLE, VSX and ASAS-SN.
    
    2MASS~07034393$+$3341346 was classified as an eclipsing binary by CRTS (\citealt{Drake2014}) and by ASAS-SN. The phased light curve is, however, peculiar, with one nearly symmetric peak and a second, higher, asymmetric peak, likely due to a combination of ellipsoidal effects and star spots.
    
    2MASS~19335971-6806348 has a clear periodic pattern, but with a spread out eclipse. Users classified this as an unknown variable, VSX as a $\delta$ Cepheid variable, and ASAS-SN as a rotational variable. 
    
    2MASS~10123394-6410057 appears to have some underlying rotational modulation with eclipses that get deeper with time. It is classified as a pulsating variable by users and a Cepheid by Gaia, but as a long period variable in VSX and a rotational variable by ASAS-SN. 

    Variable 2MASS~14015543$-$7513336 is a hot sub-dwarf star (\citealt{Geier2017}) identified as a variable star by \cite{Jayasinghe_2018}. It is a $0.132$~day detached eclipsing binary which appears to be an example of a reflection effect system. Two sectors of TESS observations are analyzed in \cite{Baran2021}. It was classified as a pulsating variable by our users and as a Cepheid by Gaia. The curious property of the full ASAS-SN phased light curve shown in Figure \ref{fig:interesting_vars_2} is that the eclipses, while clearly present, are ``filled in.'' This should not be a timing problem despite the short period, since the HJDs, while calculated for the ASAS-SN field centers, should be accurate to better than a minute while the eclipse width is roughly 20 minutes. Since individual eclipses are poorly sampled, it would be difficult to measure eclipse timing variations. 
    
    2MASS~11100731-5344255 shows no clear periodicity when phased by any period returned by the GLS periodogram, though VSX offers a period of 9.85 days. It was classified as an eclipsing binary by users, but as a rotational variable by VSX and ASAS-SN.

\section{CONCLUSION}
    We present the second data release of Citizen ASAS-SN. We started with a sample of 1665695 classifications made by a total of 7458 users of 244599 variables. There were four different groups of candidates with criterion designed to test aspects of either candidate selection or user performance. The Relaxed group, made up of noisier light curves, showed that relaxing the selection criteria used in our earlier \cite{Christy_2022_DR1} study resulted in a high junk classification rate. The Standard Group was selected as in \cite{Christy_2022_DR1}. The NP Group consisted of variables selected without computing periodograms. Their classification results were similar to the Standard Group's results, which means that it is possible to select variable candidates first and only then compute periodograms. The Machine Learning Group, made up of variables deemed real by the RF classifier, showed that the RF classifier had a false positive rate of only 2-3\%. The groups are further described in Table \ref{Table: Group}. We identified 4432 new variables, including 841 pulsating variables, 2995 rotational variables, 350 eclipsing binaries, and 246 unknown variables. 
    
    As in \cite{Christy_2022_DR1}, we find that users are able to classify eclipsing binaries with $\sim$75\% accuracy, pulsating variables with $\sim$ 73\% accuracy, and rotational variables with 50\% accuracy. Users had the most difficult time with the rotational and irregular variable stars. The low yield of new variables suggests that the variable star samples accessible to ASAS-SN in terms of magnitude and amplitude are approaching 100\% completeness.

    We show eight systems individual systems because they either show unusual variability or illustrate issues that confused the users.
    Moving forward, the goal is to complete the citizen science search across the sky, particularly to build a junk sample that
    covers more circumstances and can then be used to improve our machine learning searches. With the longer time span
    of the data and the improved sampling of short time baselines due to the longitude spacing, we will also expand the search
    to shorter periods.  The existing variable searches in ASAS-SN have been limited to sources fainter than $\sim 11$~mag
    because of saturation, but we are now able to obtain reasonably good light curves of even very saturated light curves
    using a neural network \citep{GuyAndKochanek}, which will enable us to do a systematic search reaching stars visible
    to the naked eye. We will also take closer looks at the interesting variables that users flagged, as these stars are what often give us insight into new phenomena. We will analyze user classifications once all upload batches have been fully retired. We will also look into alternative classification workflows for specific variable classes.
    
\section*{Acknowledgements}

    We thank the Zooniverse team and each volunteer who participated in Citizen ASAS-SN. We thank the Las Cumbres Observatory and their staff for its continuing support of the ASAS-SN project. We also thank the Ohio State University College of Arts and Sciences Technology Services for helping us set up and maintain the ASAS-SN variable stars database.
    
    ASAS-SN is funded  by Gordon and Betty Moore Foundation grants GBMF5490 and GBMF10501, and the Alfred P. Sloan Foundation grant G-2021-14192. Development of ASAS-SN has been supported by NSF grant AST-0908816, the Mt. Cuba Astronomical Foundation, the Center for Cosmology and AstroParticle Physics at the Ohio State University, the Chinese Academy of Sciences South America Center for Astronomy (CASSACA), the Villum Foundation, and George Skestos.
    
    CSK and KZS are supported by NSF grants AST-1907570, 2307385, and 2407206. CTC is supported by the NSF through award SOSPA9-007 from the NRAO. DMR is supported by the OSU Presidential Fellowship. The Shappee group at the University of Hawai'i is supported with funds from NSF (grants AST-1908952, AST-1911074, \& AST-1920392) and NASA (grants HST-GO-17087, 80NSSC24K0521, 80NSSC24K0490, 80NSSC24K0508, 80NSSC23K0058, \& 80NSSC23K1431). TAT is supported in part by Scialog Scholar grant 24216 from the Research Corporation. JLP acknowledges support from ANID, Millennium Science Initiative, AIM23-0001.

    This work has made use of data from the European Space Agency (ESA) mission Gaia (\url{https://www.cosmos.esa.int/ gaia}), processed by the Gaia Data Processing and Analysis Consortium (DPAC, \url{https://www.cosmos.esa.int/web/gaia/dpac/ consortium}). Funding for the DPAC has been provided by national institutions, in particular the institutions participating in the Gaia Multilateral Agreement.
    This research has made use of the VizieR catalogue access tool, CDS, Strasbourg, France. The original description of the VizieR service was published in A\&AS 143, 23.
    This research made use of Astropy, a community-developed core Python package for Astronomy (Astropy Collaboration, 2013).

\vspace{0.05cm}
\section*{Data Availability}
    The variables are publicly cataloged with the AAVSO and the ASAS-SN light curves can be obtained using the ASAS-SN Sky Patrol (\url{https://asas-sn.osu.edu}). The catalog of variables and the associated light curves are available on the ASAS-SN variable stars database (\url{https:// asas-sn.osu.edu/variables}). The external photometric data underlying this article were accessed from sources in the public domain: Gaia (\url{https://www.cosmos.esa.int/gaia}), 2MASS (\url{https://old.ipac.caltech.edu/2mass/overview/access.html}), AllWISE (\url{http://wise2.ipac.caltech.edu/ docs/release/allwise/}) and GALEX (\url{https://archive. stsci.edu/missions-and-data/galex-1/}).


\bibliographystyle{mnras}
\bibliography{main.bib}


\vspace{2cm}
\end{document}